\documentclass{ws-mpla}
\usepackage[super,compress]{cite}
\usepackage{epsfig}
\usepackage{wasysym}
\usepackage{amssymb}
\usepackage{amsmath}
\usepackage[figuresright]{rotating}
\usepackage{xcolor}

\begin{document}

\markboth{M. Ridwan and T. Mart}{Note on the electromagnetic radius of proton}

\catchline{}{}{}{}{}

\title{Note on the electromagnetic radius of proton}

\author{M. Ridwan and T. Mart}

\address{Departemen Fisika, FMIPA, Universitas Indonesia, Depok 16424, Indonesia\\
terry.mart@sci.ui.ac.id}

\maketitle

\begin{history}
\received{Day Month Year}
\revised{Day Month Year}
\end{history}

\begin{abstract}
We have analyzed the proton form factor data by using a number of 
phenomenological parameterizations (models) and extracting the proton electric
and magnetic radii. To this end we performed a global fit to all available
form factor data, with the virtual photon momentum squared $Q^2$ 
from $0.0002$ to nearly 10 GeV$^2$ for electric form factor
and from $0.015$ to 31 GeV$^2$ for magnetic one. 
Special attention was given to 
the small structure shown by the form factor data near 
$Q^2 = 0.2$ GeV$^2$. It was found that different models yield 
different structures with different numbers of minimum 
at this kinematics. Since the slope of form factor
in the limit of $Q^2\to 0$ is influenced by this structure, the extracted
proton radii are consequently different for different models. 
Our finding recommends that future experiments should focus on this 
kinematics instead of low $Q^2$. Experimental data with 
accuracies comparable to those of the latest data at low $Q^2$ would 
clearly help to clarify the effect of this structure on the 
proton charge radius. Interestingly, most of the extracted proton 
charge radii were found to be closer to the value obtained 
from the muonic hydrogen atom spectroscopy.
\end{abstract} 

\keywords{Proton; neutron; form factors; charge radius; magnetic radius.}

\ccode{PACS numbers: 13.40.Gp, 13.85.Dz, 14.20.Dh}

\maketitle

\section{Introduction}
\label{sec:introduction}
Since the work of Nobel laureate Robert Hofstadter in fifties \cite{hofstadter}, 
the elastic electron-proton scattering has become a standard tool to determine 
the radius of proton. It was found that the structure of the proton can be best
explained by using a dipole form factor with a cut-off $\Lambda^2=0.71$ GeV$^2$
and, surprisingly, in a wide range of virtual photon momentum squared $Q^2$.
Decades after that, however, it was found that this parameterization cannot explain
newer data produced by modern electron accelerator laboratories with
unprecedented precise particle detectors. Nevertheless, the dipole form 
factor $G_{\rm SD}=(1+Q^2/0.71~{\rm GeV}^2)^{-2}$ is still 
used in textbooks and state-of-the-art analyses as a standard one and 
known as the standard dipole form factor.

For present purpose, it is important to mention a number of modern 
electron-proton scattering experiments performed in the last decade. 
The first precise experiment\cite{bernauer-2010} was done in MAMI at 
Mainz in 2010. From this experiment thousands cross section data points
were extracted and fitted to obtain the electric $G_{E,p}$ and magnetic
$G_{M,p}$ form factors of proton. The proton charge radius 
$r_E=0.879\pm 0.008$ fm and the magnetic one 
$r_M=0.777\pm 0.017$ fm were obtained from these data. 
This experiment was followed by another one, using the same spectrometer 
but different technique designed to reach very small $Q^2$, i.e., the 
initial-state radiation (ISR). By using the ISR technique, the cross 
section data were obtained with $Q^2$ from 0.001 to 
0.004 GeV$^2$. The first extracted radius\cite{Mihovilovic:2017} was 
$r_E=0.810\pm 0.082$ fm. Three years later, by improving the calculation
and reinterpretation of these data, the experimentalists in Mainz 
obtained \cite{Mihovilovic:2019jiz} a radius of $r_E=0.878\pm 0.033$ fm. 

By measuring the polarization transfer in electron-proton elastic scattering
the experimentalists in Hall A of Jefferson Lab were able to determine the
ratio $\mu_pG_{E,p}/G_{M,p}$ with high precision. A global 
fit\cite{Arrington:2007ux} to the combination of these data with 
previous available data yields a proton radius of $r_E=0.875\pm 0.010$ fm.
Note that, except the previous Mainz ISR report, all results from elastic 
electron-proton scattering experiments are consistent with the 
value obtained from the Mainz-2010 one\cite{bernauer-2010}.

Very recently, the PRad collaboration\cite{Xiong:2019umf} performed another 
precise elastic electron-proton scattering experiment at Jefferson Lab. 
To reach significantly 
low $Q^2$ values, the PRad experiment used two-dimensional high-resolution 
electromagnetic calorimeter, instead of a conventional magnetic spectrometer. 
By exploiting this calorimeter, smaller scattering angles could be 
covered and, as a consequence, $Q^2$ range from $0.0002$ to 
$0.0600$ GeV$^2$ could be reached. The extracted 
proton radius was found to be $r_E=0.831\pm 0.014$ fm, which is 
surprisingly inconsistent with the values obtained from the
conventional elastic electron-proton scattering.

On the other hand, the electromagnetic radius of proton can be also determined 
by means of the spectroscopy of electronic and muonic hydrogen atoms. This is
possible since the effect of proton's finite size in hydrogen atom on hydrogen
energy\cite{miller:2019} is proportional to the proton's squared radius 
$\langle r_{E,p}^2\rangle \equiv r_E^2$. 
The first modern muonic hydrogen spectroscopic 
experiment to this end was performed at Paul Scherrer Institute by measuring
the frequency of transition between $2S_{1/2}^{F=1}$ and $2P_{1/2}^{F=2}$
states\cite{Pohl:2010zza}. The obtained proton charge radius is 
$0.84184\pm 0.00067$ fm, significantly smaller than that obtained from 
electron-proton scattering experiments. This 
result surprised the community, because the theoretical calculation used
to extract the radius is based only on quantum electrodynamics and, 
therefore, has much less uncertainties.

The publication of the first proton radius extraction from muonic hydrogen 
atom in 2010 was followed by the publications of four similar experiments, 
but using conventional electronic hydrogen. By measuring the $2S-4P$ 
transition in electronic hydrogen atom, the experimentalists in 
Garching\cite{Beyer:2017} were able to determine the radius to 
be $r_E=0.8335\pm 0.0095$ fm, which is 
consistent with the radius obtained from muonic hydrogen atom.
Another similar experiment in Paris \cite{Fleurbaey:2018} measured
the $1S-3S$ two-photon transition and reported a radius of 
$r_E=0.877\pm 0.013$ fm, consistent with that obtained from
electron-proton scattering. Two other similar experiments found the proton 
radius to be $r_E=0.833\pm 0.010$ fm \cite{Bezginov:2019}
and $r_E=0.8482\pm 0.0038$ fm \cite{Grinin:2020}, which are closer to
the result of muonic hydrogen atom.

The large differences among the extracted proton radii obtained by different
experiments sparked the so called proton radius puzzle in the community. 
Actually, before this puzzle attracted the attention of the community, 
there had been two other puzzles, i.e., the proton mass puzzle and the 
proton spin crisis. 

There have been considerable efforts devoted to resolve the proton radius 
puzzle, spanning from field theory to extra dimension. Since it is not our 
intention to discuss this puzzle, along with the proposed solutions, in 
detail, we refer the interested reader to a recent review
by Gao and Vanderhaeghen \cite{Gao_Vanderhaeghen:2022} on this topic 
for a more comprehensive information. Nevertheless, it is important 
to note that this review concludes that the proton charge radius puzzle 
has not been resolved yet. More experimental results from electron-proton 
scattering as well as from hydrogen spectroscopy are required to fully 
resolve this puzzle. A similar conclusion can be also drawn from the present
work discussed in this note.

In this note, we would like to raise the issue of high $Q^2$  proton 
form factor data and their effect on the extraction of the proton radius. 
We observe that most of experimental and theoretical analyses are focused 
on low or even very low $Q^2$ region. Of course, this is understandable, 
since by definition the radius of proton is determined by the slope of 
its form factor at $Q^2=0$. Thus, most experimental and theoretical 
investigations to this end are competing to reach the lowest possible 
$Q^2$. On the other hand, it is also clear that in the limit of 
$Q^2\to 0$ the slope of form factor would also be influenced
by the higher $Q^2$ data, if we performed a global fit (fit to all available
$Q^2$). Interestingly, in the case of proton electric form factor $G_{E,p}$
there is a minuscule structure at $Q^2\approx 0.2$ GeV$^2$,
which is almost invisible if we just plot the form factor data as a
function of $Q^2$. This structure was discussed two decades ago by 
Friedrich and Walcher and interpreted as a contribution of pion 
cloud~\cite{Friedrich:2003iz}. In this note we found that in order to 
clearly see this structure, we should present the form factors in term 
of the standard dipole one, $G_{\rm SD}$. Similar structure is also 
observed in the magnetic form factor of proton $G_{M,p}$. 

A global fit to proton form factor data has been also discussed in the 
literature, e.g., by Arrington {\it et al.} \cite{Arrington:2007ux}
and Atac {\it et al.} \cite{Atac:2021}{.}
However, comparing with the analyses of the low $Q^2$ data,
they are relatively seldom. Furthermore, these analyses usually 
did not pay a special attention on the structure existing 
at $Q^2\approx 0.2$ GeV$^2$. 
In the present note, we simultaneously 
focus on this structure and extract the proton radii from a global fit by 
using several form factor models, including the Friedrich-Walcher and 
Arrington ones. 
Similar studies focusing on the dependence of the
extracted proton radius on the form factor models and $Q^2$ cut-off can be
found, e.g., in 
Refs.~\cite{Arrington:2015yxa,Hagelstein:2018zrz,Horbatsch:2015qda,Griffioen:2015hta,Higinbotham:2015rja,Distler:2015rkm,Yan:2018bez,Alarcon:2018zbz,Barcus:2019skg,Alarcon:2020kcz,Paz:2020prs,Borisyuk:2020pxo}{.}

The organization of this note is as follows. 
In Sec.~\ref{sec:formalism} we revisit the formalism 
required for extracting the electromagnetic radius of proton. 
We present several form factor parameterizations (models), which
include the dipole, double-dipole, Friedrich-Walcher, and Arrington ones. We 
also derive the corresponding error bar formulas required for our present
analysis. In Sec.~\ref{sec:result} we present the result of our
analysis. We start with reanalyzing the PRad data and then we continue
by including higher $Q^2$ data in our analysis. The obtained proton charge
and magnetic radii in the present work will be discussed in this 
section. We close this note by presenting the summary and conclusion
in Sec.~\ref{sec:conclusion}.

\section{Formalism}
\label{sec:formalism}

\subsection{Elastic Electron-Proton Scattering}
\begin{figure}[t]
  \begin{center}
    \leavevmode
    \epsfig{figure=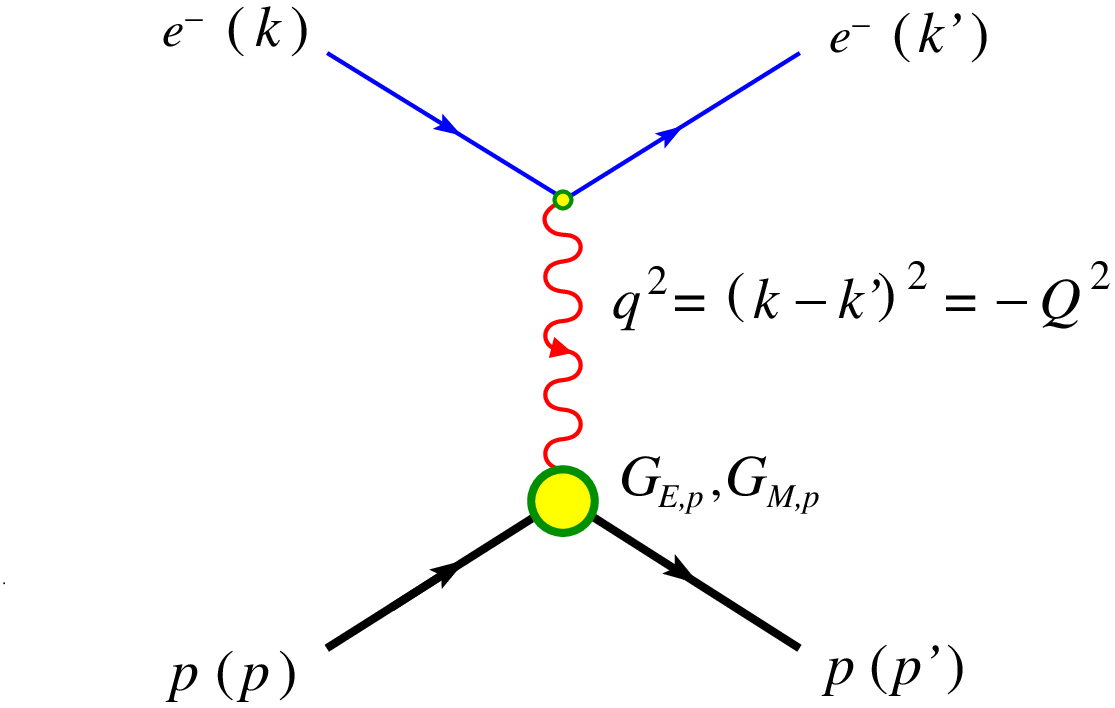,width=70mm}
    \caption{Kinematics of the electron-proton elastic scattering.
    The four-momenta of the initial and final electrons are defined
    as $k^\mu=(E,{\bf k})$ and ${k'}^\mu=(E',{\bf k}')$,
    respectively.}
   \label{fig:kinema} 
  \end{center}
\end{figure}

Conventionally, the elastic electron-proton scattering process is
depicted in Fig.~\ref{fig:kinema}, where the one-photon exchange
process is assumed. The kinematical variable required in the following
discussion is only the opposite of the virtual 
photon momentum squared $Q^2=-q^2$.
The cross section is given in most standard textbooks. It is
usually expressed in terms of the electric and magnetization (Sachs)
form factors $G_E$ and $G_M$, which depend solely on $Q^2$, i.e.,
\begin{equation}
\label{eqn:ep_cross}
\left(\frac{d\sigma}{d\Omega}\right)_N = \left(\frac{d\sigma}{d\Omega}\right)_{\rm Mott}\frac{1}{1 + \tau}\left[ G_E^2(Q^2) + \frac{\tau}{\epsilon}G_M^2(Q^2)\right] ~,
\end{equation}
where $\tau = {Q^2}/{4M^2}$, $M$ is the proton mass, and 
the Mott cross section is given by
\begin{eqnarray}
\left(\frac{d\sigma}{d\Omega}\right)_{\rm Mott} = \frac{\alpha^2 \cos^ 2({\theta}/{2})}{4E^2 \sin^4 ({\theta}/{2})} \frac{E'}{E} ~,
\end{eqnarray}
with $E$ and $E'$ the initial and final electron energies, 
respectively, defined in Fig.~\ref{fig:kinema},
whereas the virtual photon polarization reads
\begin{eqnarray}
\epsilon=\left[1+2(1+\tau)\tan^2\frac{\theta}{2}\right]^{-1} ~{.}
\end{eqnarray}
For completeness, we have also to mention that $\theta$
is the electron scattering angle and in the limit of $Q^2 \to 0$ 
the form factors are normalized to 
\begin{eqnarray}
G_E(0) = 1, \,\,\, G_M(0) = 1 + \kappa = \mu 
~~{\rm or}~~ G_M(0)/\mu = 1 ~,
\end{eqnarray}
where $\mu$ is the proton magnetic moment in the unit 
of nuclear magneton.

In the non-relativistic view it is common to interpret the
form factors $G_E$ and $G_M$ as the Fourier transforms of charge
and magnetization distributions inside the proton 
\begin{equation}
G_{E,M}(Q^2) = \int \rho_{E,M}({\bf r})e^{i{\bf q}\cdot{\bf r}}\,d{\bf r}~.
\end{equation}
However, it is
well known that this is not true in the relativistic case, where 
the relativistic wave functions are frame dependent and, therefore,
the interpretation is only valid in the case of Breit frame. 
There have been solutions proposed to solve this problem. For instance, see
Refs. \cite{Lorce:2020onh,Gao_Vanderhaeghen:2022}{.}  
Nevertheless, for the operational 
definition of the form factor slope at photon point, we can Taylor
expand the form factors as \cite{Gao_Vanderhaeghen:2022}
\begin{equation}
  \label{eqn:pers7}
  G_{E,M}(Q^2) = 1 - \frac{1}{6}\langle r_{E,M}^2\rangle\, Q^2 + 
  \frac{1}{120}\langle r_{E,M}^4\rangle\, Q^4 - \cdots ~,
\end{equation}
from which we can calculate the corresponding root mean square (rms) radius as
\begin{eqnarray}
\label{eqn:radius}
r_{E,M} = 
\sqrt{\langle r_{E,M}^2\rangle} = \left[-6\frac{dG_{E,M}(Q^2)}{dQ^2} 
\Big|_{Q^2=0}\right]^{1/2} ~.
\end{eqnarray}
It is important to emphasize here that the squared 
radius defined by
Eq.~(\ref{eqn:radius}) is, up to the factor of $-6$, nothing but 
the slope of the form factor at $Q^2=0$.
Thus, although the recent and modern data bring us much closer to this photon
point, the determination of radius by using this method is merely an
estimate, which is strongly influenced by the data at both small and
large $Q^2$. The latter is the main interest in this note and will
be shown in the next section, when we discuss the result of our analysis.

\subsection{Phenomenological Fits of the Form Factors}
For decades the standard dipole form factor 
\begin{equation}
\label{eq:std}
G_{\rm SD}(Q^2) = \left(1 + \frac{Q^2}{0.71~{\rm GeV}^2}\right)^{-2} 
= \left[1 + \frac{Q^2}{(0.84~{\rm GeV})^2}\right]^{-2}
\end{equation}
has been used to fit the available experimental data, from very low 
to very high $Q^2$. The impressive agreement between Eq.~(\ref{eq:std})
and experimental data has led to the question whether the special
parameter $0.71$ GeV$^2$ has a physical meaning related to the structure
of the proton. Unfortunately, there is none \cite{Friedrich:2003iz}.
Moreover, it was soon realized that the $Q^2$ distribution of 
experimental data are not smooth; they form a structure at 
$Q^2\approx 0.2$ GeV$^2$, which is clearly not reproducible by
using Eq.~(\ref{eq:std}). Nevertheless, it is always interesting
to use the dipole form factor
\begin{equation}
\label{eq:dipole}
G_{\rm D}(Q^2) = \left(1 + \frac{Q^2}{\Lambda^2}\right)^{-2} ~,
\end{equation}
in the study of general baryon form factors and to fit
the parameter $\Lambda$, since it is the simplest possible form factor 
to this end. Furthermore, we are also curios to investigate the limit
of dipole form factor in view of the recent available experimental data
in the low $Q^2$ region. Note that by using Eq.~(\ref{eqn:radius})
to the dipole parameterization given by Eq.~(\ref{eq:dipole}) we obtain
the rms radius of
\begin{equation}
  \label{eq:rad_dipole}
  r_{\rm D} \equiv \sqrt{\langle r_{\rm D}^2 \rangle} 
  = \frac{\sqrt{12}}{\Lambda} ~,
\end{equation}
with error given by
\begin{equation}
  \label{eq:error_rad_dipole}
  \Delta r_{\rm D}  = r_{\rm D}\, \frac{\Delta\Lambda}{\Lambda} ~.
\end{equation}
Note that using the form factor parameter given by the standard dipole
of Eq.~(\ref{eq:std}), $\Lambda=0.84~{\rm GeV}=4.26~{\rm fm}^{-1}$, the
corresponding radius according to Eq.~(\ref{eq:rad_dipole}) is 0.81 fm,
much smaller than that obtained from present experiments using 
elastic electron-proton scattering.

A modification to the dipole form factor, Eq.~(\ref{eq:dipole}), 
is known as the double dipole (DD) one. 
The DD form factor is certainly more flexible than the dipole one due
to the addition of two new parameters $a_0$ and $a_2$,
whereas $a_1\equiv\Lambda$, i.e.,
\begin{equation}
\label{eqn:double_dipole}
G_{\rm DD} (Q^2,a_{0},a_{1},a_{2}) = a_{0}\left(1 + \frac{Q^2}{a_{1}}
\right)^{-2}+(1-a_{0})\left(1 + \frac{Q^2}{a_{2}}\right)^{-2} ~.
\end{equation}
Note that $G_{\rm DD}$ is properly normalized at photon point, i.e.,
$G_{\rm DD}(0,a_{0},a_{1},a_{2}) = 1$.
It is straightforward to calculate the rms radius given by
Eq.~(\ref{eqn:double_dipole}). The obtained radius reads
\begin{equation}
  \label{eq:rad_doub_dipole}
  r_{\rm DD}  = \sqrt{12}
  \left( \frac{a_0}{a_1}+\frac{1-a_0}{a_2} \right)^{1/2}~,
\end{equation}
with the corresponding error 
\begin{equation}
  \label{eq:error_rad_doub_dipole}
  \Delta r_{\rm DD}  = \frac{6}{r}\, \left\{
    (\Delta a_0)^2\left(\frac{1}{a_1}-\frac{1}{a_2} \right)^2 +
    (\Delta a_1)^2\left(\frac{a_0}{a_1^2} \right)^2 +
    (\Delta a_2)^2\left(\frac{1-a_0}{a_2^2} \right)^2
    \right\}^{1/2} ~.
\end{equation}

To account for the small structure at $Q^2\approx 0.2$ GeV$^2$
Friedrich and Walcher (FW) added a non-smooth term $G_{\rm ns}$
to the DD form factor given by Eq.~(\ref{eqn:double_dipole}), with
\begin{equation}
\label{eqn:non-smooth_FW}
G_{\rm ns} (Q^2,Q_b,\sigma_{b})= \exp{\left\{-\frac{1}{2}
    \left(\frac{Q-Q_b}{\sigma_{b}}\right)^2\right\}}  + \exp{\left\{
    -\frac{1}{2}\left(\frac{Q+Q_b}{\sigma_{b}}\right)^2\right\}} ~.
\end{equation}
Thus, the complete expression for the FW form factor is 
\cite{Friedrich:2003iz}
\begin{eqnarray}
\label{eqn:FF_FW}
G_{\rm FW} (Q^2,a_{0},a_{1},a_{2},Q_b,\sigma_{b})
= G_{\rm DD} (Q^2,a_{0},a_{1},a_{2}) 
+ a_b\, Q^2\, G_{\rm ns} (Q^2,Q_b,\sigma_{b}) ~.
\end{eqnarray}
It is obvious that the corresponding expressions of radius and 
its error can be obtained from Eqs.~(\ref{eq:rad_doub_dipole}) 
and (\ref{eq:error_rad_doub_dipole}) with the addition of the
non-smooth contribution. Explicitly, they read
\begin{equation}
  \label{eq:rad_FW}
  r_{\rm FW}  = \sqrt{12}\left\{ 
    \frac{a_0}{a_1}+\frac{1-a_0}{a_2} 
    -a_b\, \exp\left(-\frac{Q_b^2}{2\sigma_b^2}\right)
  \right\}^{1/2}~,
\end{equation}
and
\begin{eqnarray}
  \label{eq:error_rad_FW}
  \Delta r_{\rm FW}  &=& \frac{6}{r}\, \left[
    (\Delta a_0)^2\left(\frac{1}{a_1}-\frac{1}{a_2} \right)^2 +
    (\Delta a_1)^2\left(\frac{a_0}{a_1^2} \right)^2 +
    (\Delta a_2)^2\left(\frac{1-a_0}{a_2^2} \right)^2 \right.
     \nonumber\\ &&
    \left. +\left\{ 
      (\Delta a_b)^2 + (\Delta Q_b)^2\,\frac{a_b^2\, Q_b^2}{\sigma_b^4}
      + (\Delta \sigma_b)^2\,\frac{a_b^2\, Q_b^4}{\sigma_b^6}
    \right\}\, \exp\left(-\frac{Q_b^2}{\sigma_b^2}\right)
    \right]^{1/2} .
\end{eqnarray}
The last model investigated in this work is obtained from the
Arrington {\it et al.} ansatz \cite{Arrington:2007ux}. In this
ansatz the form factor is expanded in a polynomial form, i.e.,
\begin{equation}
  \label{eq:arrington} 
  G_{\rm Arr} = \frac{1 + \sum_{i=1}^{n}a_i \tau^i}{1 + 
    \sum_{i =1}^{n+2}b_i \tau^i} ~,
\end{equation}
with $\tau = {Q^2}/{4M^2}$ and $M$ is the proton mass. 
Although the summation in Eq.\,(\ref{eq:arrington}) 
could be performed up to infinity, Arrington {\it et al.} 
limited it to $n = 3$. Thus, Eq.~(\ref{eq:arrington}) reduces to
\begin{eqnarray}
  \label{eq:arrington1} 
  G_{\rm Arr} = \frac{1 + a_1 \tau + a_2 \tau^2 + a_3 \tau^3}{1 + b_1 \tau + b_2 \tau^2 + b_3 \tau^3 + b_4 \tau^4 + b_5 \tau^5} ~.
\end{eqnarray}
Since only parameters $a_1$ and $b_1$ are proportional to the $Q^2$,
only these parameters survive in the limit of photon point.
As a consequence, 
the corresponding radius and error are given only
in terms of $a_1$ and $b_1$.
Explicitly, they are given by
\begin{equation}
  \label{eq:rad_arrington}
  r_{\rm Arr}  = \sqrt{\frac{3}{2}}\,
  \frac{\left( b_1-a_1 \right)^{1/2}}{M}~,
\end{equation}
with the corresponding error 
\begin{equation}
  \label{eq:error_arrington}
  \Delta r_{\rm Arr}  = \frac{3}{4M^2r_{\rm Arr}}\, \left\{
    (\Delta a_1)^2 + (\Delta b_1)^2
    \right\}^{1/2} ~.
\end{equation}

\section{Results and Discussion}
\label{sec:result}
The minimization process in this work was performed by using the 
{\small MINUIT} code \cite{James:1975dr}. We combined the 
{\small SIMPLEX} and {\small MIGRAD} minimizers in the 
minimization process until the {\small MIGRAD} reaches the 
convergence. All errors reported here are only those 
obtained from the {\small MINUIT} output.

\subsection{Revisiting PRad Data}
\begin{figure}[t]
  \begin{center}
    \leavevmode
    \epsfig{figure=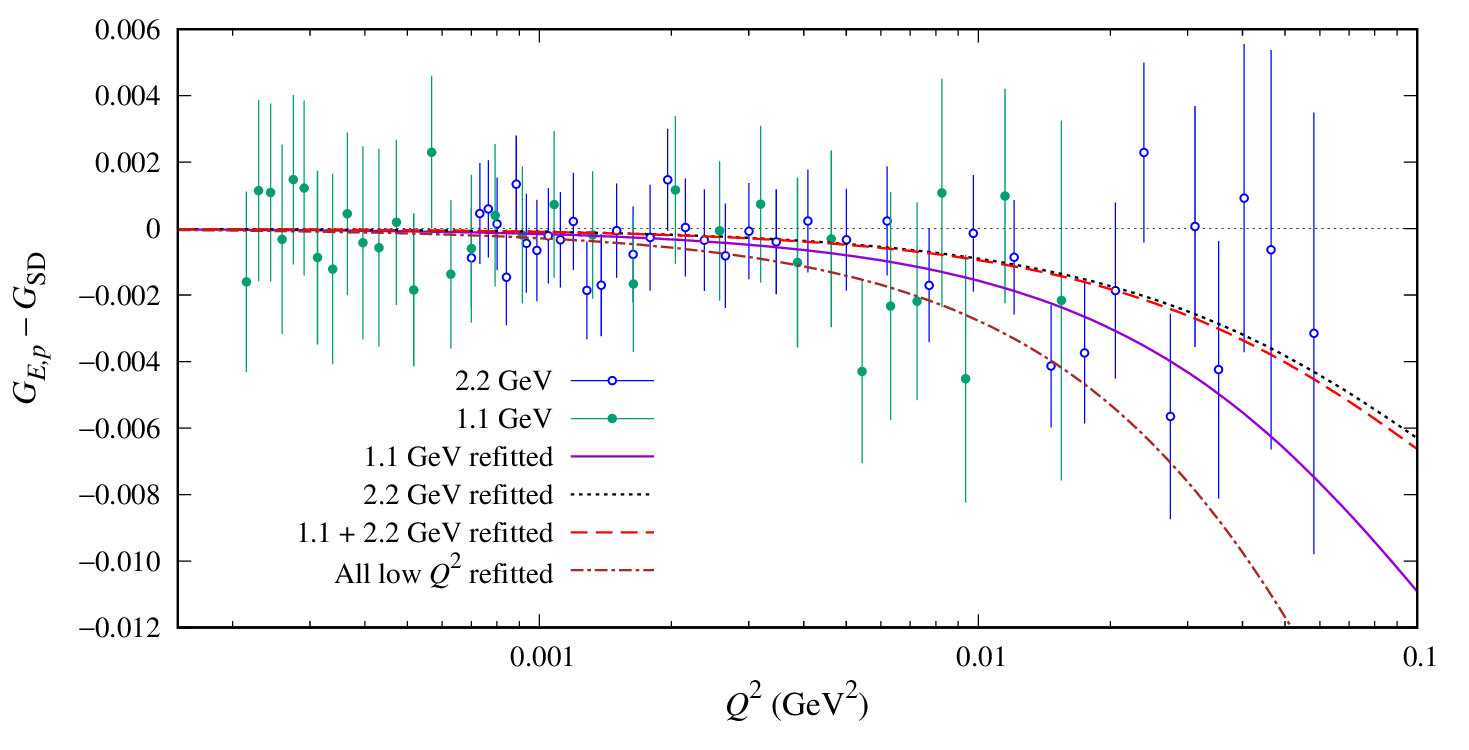,width=130mm}
    \caption{The relative electric form factor of proton to the standard
      dipole one, $G_{E,p}-G_{\rm SD}$, 
      measured by the PRad collaboration~\cite{Xiong:2019umf} with electron
      beam energies $E=1.1$ (closed circles) and $2.2$ (open circles) GeV,
      compared with the results of fitting to different sets of experimental 
      data.
      }
   \label{fig:electric_prad} 
  \end{center}
\end{figure}

As stated in the Introduction, the latest measurement by the PRad collaboration
\cite{Xiong:2019umf} is very interesting to analyze, since the corresponding 
data yield a smaller proton radius, compared to those obtained by different
collaborations. Furthermore, the extracted radius is consistent with the 
muonic-hydrogen atom experiment~\cite{Pohl:2010zza}. It is notoriously 
known that the data points with
very small $Q^2$ have the form factors very close to one and, therefore, 
difficult to distinguish.
To overcome this problem, in Fig.~\ref{fig:electric_prad} we plot the relative
electric form factor of proton to the standard dipole one, $G_{E,p}-G_{\rm SD}$.
As will be explained in the next figures, for this purpose the use of 
$G_{E,p}-G_{\rm SD}$ is superior than that of $G_{E,p}/G_{\rm SD}$ or
$G_{E,p}$ itself. Furthermore, for the sake of brevity, let us define 
$\Delta\equiv G_{E,p}-G_{\rm SD}$.

At first glance we may say that both the 1.1 and 2.2 GeV PRad data are 
randomly scattered around the standard dipole $G_{\rm SD}$ curve, i.
e., around $\Delta=0$. Only 
for $Q^2\ge 0.005$ GeV$^2$ more data start to deviate from the standard dipole
parameterization with negative $\Delta$, albeit with relatively large error
bars. 

In their report, PRad collaboration used the parameterization similar to the
Arrington model, i.e., Eq.~(\ref{eq:arrington}) with $n=1$ both 
in the numerator
and denominator. Furthermore, a more complicated treatment was also used in order
to achieve an optimal $\chi^2$. In this analysis, we will only use the dipole form
factor, since it is simpler and more under-control. The result of fitting this
dipole form factor to different data sets is shown in Fig.~\ref{fig:electric_prad},
where for comparison 
we also show the result of fitting to all available low $Q^2$ data that include
older experiments.

It is apparent from Fig.~\ref{fig:electric_prad} that the 1.1 GeV data set yields
slightly larger slope, i.e., larger radius and smaller $\Lambda$. The more scattered
data points of the 2.2 GeV data set, especially for $Q^2\ge 0.005$ GeV$^2$, 
tend to decrease the extracted radius. Thus, different data sets from the same
experiment can produce different proton radii. Furthermore, since fitting both 
1.1 and 2.2 GeV data sets simultaneously yields almost similar result to 
fitting to 2.2 GeV data set 
(compare the dotted and dashed curves in Fig.~\ref{fig:electric_prad}), this
indicates that the data with higher $Q^2$ values are more decisive for 
determination of the proton radius. This is obvious, since the slope is
strongly dependent on the higher $Q^2$ behavior of form factor. 

\begin{figure}[t]
  \begin{center}
    \leavevmode
    \epsfig{figure=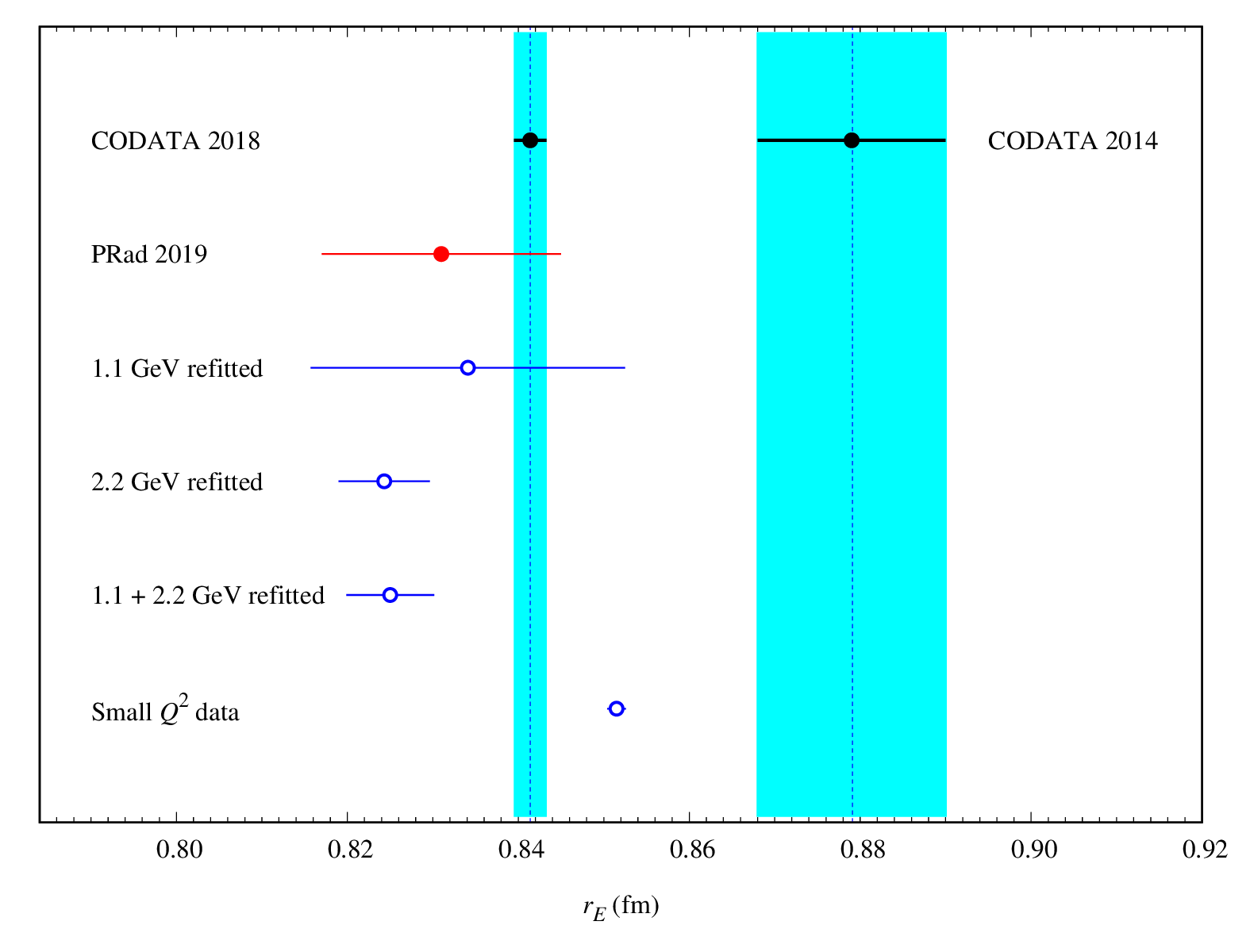,width=130mm}
    \caption{The extracted radii of proton obtained from fitting to different
      experimental databases shown in Fig.~\ref{fig:electric_prad}, compared
      with those reported by CODATA 2014 \cite{Mohr:2015ccw} and CODATA 2018
      \cite{Tiesinga:2021myr}.
      }
   \label{fig:radius_prad} 
  \end{center}
\end{figure}

The extracted radii obtained from different data sets given in 
Fig.~\ref{fig:electric_prad} are shown in Fig.~\ref{fig:radius_prad}, 
where we also show the original value extracted by the PRad collaboration 
and the recommended values from CODATA 2014 \cite{Mohr:2015ccw} 
and CODATA 2018 \cite{Tiesinga:2021myr}. As originally 
reported by the PRad collaboration~\cite{Xiong:2019umf}, they obtained 
the radius consistent with the later CODATA value. As shown in 
Fig.~\ref{fig:radius_prad}, in spite of the use of different 
parameterization, within the obtained error bars all values extracted 
in the present work by using different PRad data sets are consistent 
with the original one. The most consistent radius is obtained from 
fitting the 1.1 GeV data. The use of 2.2 GeV data (individually and 
simultaneously) yields smaller values and error bars. This is in line with
the previous finding in the discussion of Fig.~\ref{fig:electric_prad}.
Therefore, we confirm the finding of the PRad collaboration that their data
lead to smaller radius, closer to the value obtained 
from the muonic hydrogen atom spectroscopy. 

We also note that fitting to all other data with small $Q^2$, i.e., up to 
0.06 GeV$^2$, results in a larger radius. Nevertheless, it is still 
closer to the value from muonic hydrogen spectroscopic measurement.
This is caused by the fact that
there are data smaller than the PRad data in the 
$0.01\lesssim Q^2\lesssim 0.05$ GeV$^2$ region, which tend to
shift the $\Delta$ curve down, increasing the slope and, therefore, the
extracted radius.
We have also investigated the effect of the PRad data on this fitting. We found
that the effect is small, the form factor cut-off slightly increases from
$0.8012\pm 0.0010$ to $0.8024\pm 0.0010$ GeV after the inclusion of
the PRad data. The small error indicates that the dipole form factor has
very small degree of freedom if we include all available data. This is understood
since the number of free parameter is only one. As we will see later, this is
not the case, when we use different form factors with more free parameters.

\subsection{Fitting to Higher $Q^2$ Data}
Having shown that the higher $Q^2$ data may significantly influence the result
of extracting the proton charge radius at the $Q^2\approx 0$ region, we are 
in the position to investigate the effect of higher $Q^2$ data to this end. 
As stated in Sec.~\ref{sec:formalism}, there is a visible structure at 
$Q^2\approx 0.2$ GeV$^2$ which forces us to introduce some physics to account this 
phenomenon. Obviously, the use of dipole form factor is inadequate to fit this
structure. An alternative solution has been proposed by Friedrich and 
Walcher~\cite{Friedrich:2003iz}. 

\label{append:FFdata}
\begin{table}[!]
	\tbl{Experimental data used 
          in the present analysis along with their references
          and corresponding virtual photon momentum squared 
          ($Q^2$) ranges.
          }
          {\begin{tabular}{@{}lcc@{}} \toprule
		Form Factor & 
		 $Q^2$ Range (GeV$^2$) & Reference  \\
		\hline\\[-2ex]
		$G_{E,p}$ 
		& 0.78 - 1.75
		& Albrecht \textit{et al.}~\cite{Albrecht:1965ki} \\
		& 1.75 - 8.83
		& Andivahis \textit{et al.}~\cite{Andivahis:1994rq} \\
		& 0.39 - 4.095
		& Bartel \textit{et al.}~\cite{Bartel:1966zza} \\
		& 0.39 - 4.09
		& Behrend \textit{et al.}~\cite{Behrend:1966cg} \\
		& 0.39 - 1.95
		& Berger \textit{et al.}~\cite{Berger:1971kr}\\
		& 0.975 - 1.755
		& Berkelman \textit{et al.}~\cite{Berkelman:1963zz}\\
		& 0.0152 - 0.5524
		& Bernauer \textit{et al.} \cite{Bernauer:2010zga} \\
		& 0.00136 - 0.123
		& Borkowski \textit{et al.}~\cite{Borkowski:1975ume} \\
                & 0.389 - 1.752
		& Chen \textit{et al.}~\cite{Chen:1966zz} \\
		& 0.65 - 5.2
		& Christy \textit{et al.}~\cite{Christy:2004rc} \\
		& 0.273 - 1.755
		& Hanson \textit{et al.}~\cite{Hanson:1973vf}\\
		& 0.0058 - 5.2
		& Hohler \textit{et al.}~\cite{Hohler:1976ax} \\
		& 0.156 - 0.858
		& Janssens \textit{et al.}~\cite{Janssens:1965kd}\\
		& 0.00585 - 0.3097
		& Murphy \textit{et al.}~\cite{Murphy:1974zz} \\
		& 0.0389 - 1.75
		& Price \textit{et al.}~\cite{Price:1971zk} \\
		& 0.00546 - 0.0546
		& Simon \textit{et al.}~\cite{Simon:1980hu}\\
		& 0.15 - 9.7
		& Walker \textit{et al.}~\cite{Walker:1993vj}\\
		& 0.00134 - 0.0154
		& Weber \textit{et al.}~\cite{Weber:2017dph} \\
		& 0.000215 - 0.01547
		& Xiong \textit{et al.}~\cite{Xiong:2020kds} \\
		& 0.0007 - 0.05819
		& Xiong \textit{et al.}~\cite{Xiong:2020kds} \\
		\hline \\[-2ex]
		$G_{M,p}/\mu_p$ 
		& 0.78 - 1.75
		& Albrecht \textit{et al.}~\cite{Albrecht:1965ki} \\
		& 1.75 - 8.83
		& Andivahis \textit{et al.}~\cite{Andivahis:1994rq}\\
		& 2.883 - 31.28
		& Arnold \textit{et al.}~\cite{Arnold:1986nq} \\
		& 0.39 - 4.095
		& Bartel \textit{et al.}~\cite{Bartel:1966zza} \\
		& 0.67 - 3.0
		& Bartel \textit{et al.}~\cite{Bartel:1973rf} \\
		& 0.39 - 4.28
		& Behrend \textit{et al.}~\cite{Behrend:1966cg} \\
		& 0.39 - 1.95
		& Berger \textit{et al.}~\cite{Berger:1971kr} \\
		& 0.975 - 1.755
		& Berkelman \textit{et al.}~\cite{Berkelman:1963zz} \\
		& 0.0152 - 0.5524
		& Bernauer \textit{et al.}~\cite{Bernauer:2010zga} \\
		& 0.021 - 0.123
		& Borkowski \textit{et al.}~\cite{Borkowski:1975ume} \\
		& 0.49 - 1.75
		& Bosted \textit{et al.}~\cite{Bosted:1990hy} \\
		& 0.389 - 3.89
		& Chen \textit{et al.}~\cite{Chen:1966zz} \\
		& 0.65 - 5.2
		& Christy \textit{et al.}~\cite{Christy:2004rc} \\
		& 0.669 - 25.03
		& Coward \textit{et al.}~\cite{Coward:1967au}\\
		& 0.273 - 1.755
		& Hanson \textit{et al.}~\cite{Hanson:1973vf} \\
		& 0.017 - 5.2
		& Hohler \textit{et al.}~\cite{Hohler:1976ax} \\
		& 0.156 - 1.17
		& Janssens \textit{et al.}~\cite{Janssens:1965kd} \\
                & 0.999 - 25.03
		& Kirk \textit{et al.}~\cite{Kirk:1972xm} \\
		& 1.858 - 15.754
		& Longwu Ou \textit{et al.}~\cite{Ou:2019ing} \\
		& 0.0389 - 3.0
		& Price \textit{et al.}~\cite{Price:1971zk} \\
		& 2.862 - 31.2
		& Sill \textit{et al.}~\cite{Sill:1992qw}\\
		& 0.15 - 9.7
		& Walker \textit{et al.}~\cite{Walker:1993vj} \\
                \botrule
	\end{tabular}\label{exp_data} }
\end{table}

A list of references to the available experimental data is given in 
Table~\ref{exp_data}. From this Table
it is apparent that the high $Q^2$ data originate from relatively older
experiments. The newest experiments are racing to reach the lowest possible 
$Q^2$ region. 

The results of fitting free parameters of the dipole,
double-dipole, Friedrich-Walcher, and Arrington form factors are given in 
Tables~\ref{tab:dipole}, \ref{tab:d_dipole}, \ref{tab:FW-FF}, and
\ref{tab:Arrington-FF}, respectively. From these Tables it is obvious that
the new data deviate significantly from the standard dipole parameterization.
The dipole parameterization can give a better agreement with data, but it is
also clear that the double-dipole form factor yields a much better agreement
due to the larger number of free parameters. This is also true for the 
Friedrich-Walcher and Arrington form factors, except in these form factors
more physics is involved. Nevertheless, due to the 
large data 
errors in the large $Q^2$ region, the smaller $\chi^2$ does not directly
show the best parameterization, especially when we discuss the structure
at $Q^2\approx 0.2$ GeV$^2$.

\begin{figure}[t]
  \begin{center}
    \leavevmode
    \epsfig{figure=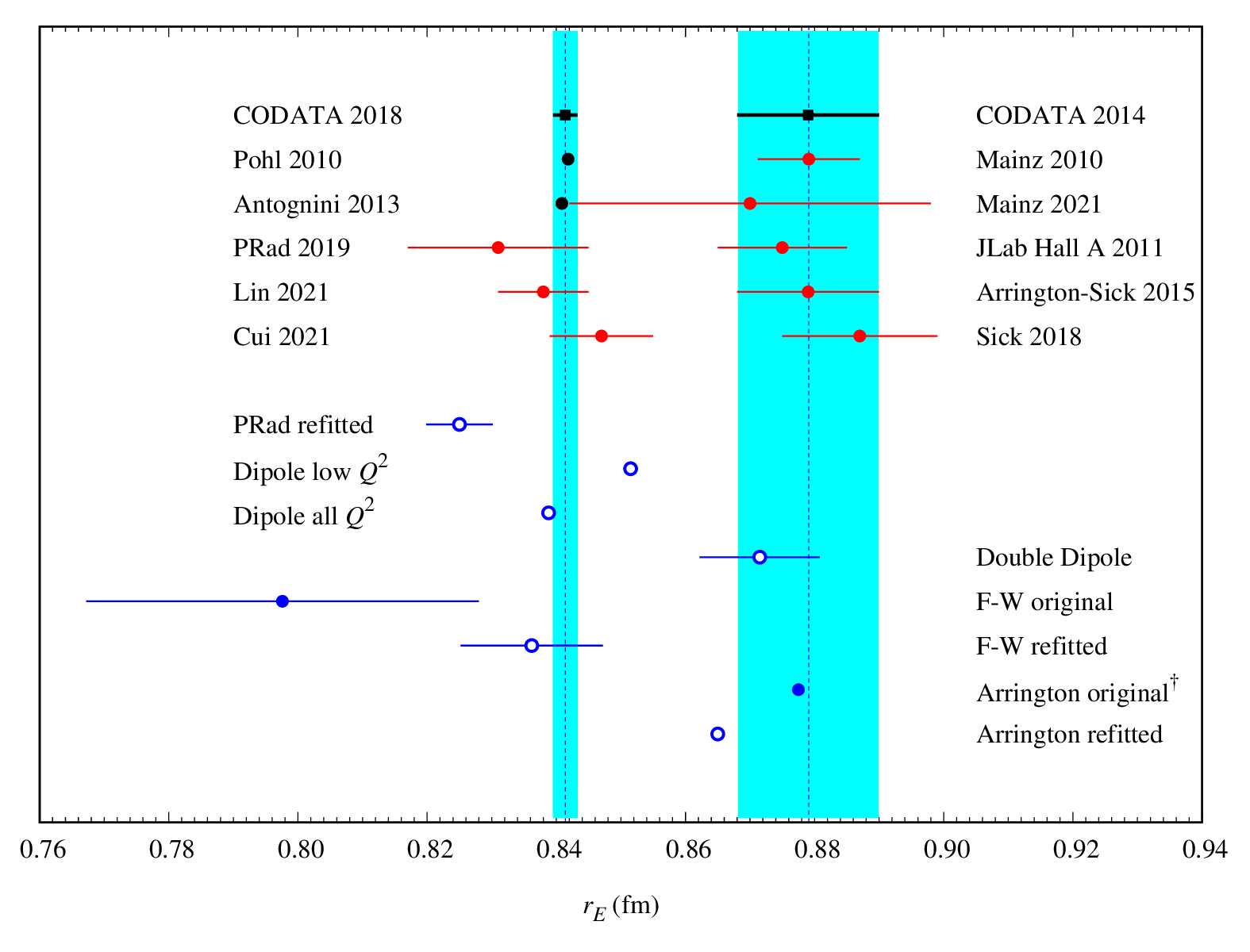,width=130mm}
    \caption{The extracted proton electric rms radius 
      $r_{E}\equiv \sqrt{\langle r_{E,p}^2 \rangle}$ obtained
      in the present and previous works 
      compared with the
      recommended values of the CODATA-2014 \cite{Mohr:2015ccw}
      and CODATA-2018 \cite{Tiesinga:2021myr}.
      The previous
      works are Pohl 2010 
      \cite{Pohl:2010zza}, Mainz 2010 \cite{bernauer-2010}, 
      Antognini 2013 \cite{Antognini:2013txn}, 
      Mainz 2021 \cite{Mihovilovic:2019jiz}, 
      PRad 2019 \cite{Xiong:2019umf},
      JLab Hall A 2011 \cite{Zhan:2011ji}, 
      Lin 2021 \cite{Lin:2021umk}, 
      Arrington-Sick 2015 \cite{Arrington:2015ria},
      Cui 2021 \cite{Cui:2021vgm}, and 
      Sick 2018 \cite{Sick:2018fzn}. The previous works
      obtained from analyses of $ep$ scattering data are
      marked with red circles, whereas those obtained from
      muonic hydrogen atom are marked with solid black ones. 
      All results obtained from present work are shown by using
      open circles. Note that the original
      Arrington model did not report error bars.}
   \label{fig:radius_el}
  \end{center}
\end{figure}

A quick glance at Tables~\ref{tab:dipole}-\ref{tab:Arrington-FF}
reveals that the extracted radii from all refitted models are smaller 
than the values obtained from the conventional elastic 
electron-proton scattering. This result is corroborated
by our result in the previous discussion on the new PRad data and shown
in Fig.~\ref{fig:radius_prad}. However, we will discuss this later, 
when we compare our present result with those given by previous works
in Fig.~\ref{fig:radius_el}, because it is easier to analyze them
by using this figure.

\begin{table}[htb] 
  \centering
  \tbl{Cut-off parameter of the standard dipole form factor used in the 
    present analysis and that of the dipole form factor extracted from all
    existing data up to 10 GeV. Note that $N_{\rm dof}=N_{\rm data}-N_{\rm par}$,
    with $N_{\rm dof}$, $N_{\rm data}$, and $N_{\rm par}$ are numbers of degrees
    of freedom, data, and parameters, respectively.
    }
          {\begin{tabular}{@{}ccc@{}} \toprule
		Parameter        & ~~~~Standard Dipole~~~~ & Dipole\\
		\hline\\[-2.5ex]
		$\Lambda$ (GeV)  & 0.8426          & $0.8145\pm 0.0003$ \\
                $r$ (fm) &$0.8109$ & $0.8388\pm 0.0003$ \\
		\hline\\[-2.5ex]
		$\chi^2$         & 12031           & 1312 \\
		$N_{\rm data}$   & 357             & 357 \\ 
           $\chi^2 /N_{\rm dof}$ & 33.701          & 3.675 \\
           \botrule
	\end{tabular}\label{tab:dipole}}
\end{table}

\begin{table}[htb] 
  \centering
  \tbl{As in Table~\ref{tab:dipole}, but for the double dipole form factor.
    }
          {\begin{tabular}{@{}cc@{}} \toprule
		Parameter        & Extracted from data \\
		\hline\\[-2.5ex]
		$a_0$ & $0.9803\pm 0.0023$ \\
		$a_1$ (GeV$^2$)& $0.6932\pm 0.0023$ \\
		$a_2$ (GeV$^2$)& $0.0925\pm 0.0117$ \\
                $r$ (fm) &$0.8714 \pm 0.0093$ \\
		\hline\\[-2.5ex]
		$\chi^2$ & 702  \\
		$N_{\rm data}$ & 357  \\ 
           $\chi^2 /N_{\rm dof}$ & 1.9674  \\
		\botrule
	\end{tabular}\label{tab:d_dipole}}
\end{table}

\begin{table}[htb] 
  \tbl{As in Table~\ref{tab:dipole}, but for the Friedrich-Walcher form 
    factor \cite{Friedrich:2003iz}.
    }
    {\begin{tabular}{@{}ccc@{}} \toprule
		Parameter        & Extracted from data & Original values\\
		\hline\\[-2.5ex]
		$a_{0}$             & $1.0031\pm 0.0016$ &$1.041\pm 0.040\,$\\
		$a_{1}$ (GeV$^2$)   & $0.6774\pm 0.0020$ &$0.765\pm 0.066\,$\\
		$a_{2}$ (GeV$^2$)   & $6.50\pm 4.43$     &$6.2\pm 5.0\,$  \\
		$a_{b}$ (GeV$^{-2}$)& $-0.1227\pm 0.0068~~$&$-0.23\pm 0.18~~$\\
		$Q_{b}$ (GeV)       & $0.1969\pm 0.0085$ &$0.07\pm 0.88$  \\
		$\sigma_{b}$ (GeV)  & $0.1001\pm 0.0811$ &$0.27\pm 0.29$  \\
                $r$ (fm)            & $0.8362\pm 0.0110$ &$0.7976\pm 0.0304$\\
		\hline\\[-2.5ex]
		$\chi^2$ & 608  & 59.71\\
		$N_{\rm data}$ & 357 & 64 \\ 
           $\chi^2 /N_{\rm dof}$ & 1.703 & 0.933 \\
		\botrule
	\end{tabular}\label{tab:FW-FF}}
\end{table}

\begin{table}[htb] 
  \centering
  \tbl{As in Table~\ref{tab:dipole}, but for the Arrington form 
    factor \cite{Arrington:2007ux}. Note that the parameter error bars in 
    the original values of Arrington model were not reported in their 
    paper \cite{Arrington:2007ux}.
    }
    {\begin{tabular}{@{}ccc@{}} \toprule
		Parameter        & Extracted from data & Original values\\
		\hline\\[-2.5ex]
		$a_{1}$ & $-2.4028\pm 0.0580~~$ &$3.439$ \\
		$a_{2}$ & $ 19.721\pm 0.7835$ &$-1.602~~$ \\
		$a_{3}$ & $ 29.998\pm 20.917$ &$0.068$ \\
		$b_{1}$ & $ 8.8857\pm 0.0630$&$15.055$ \\
		$b_{2}$ & $-4.9173\pm 4.4649~~$&$48.061$ \\
		$b_{3}$ & $ 469.26\pm 15.661$&$99.304$ \\
		$b_{4}$ & $-530.54\pm 72.354~~$&$0.012$ \\
		$b_{5}$ & $ 2807.5\pm 125.49$&$8.650$ \\
                $r$ (fm)& $0.8650\pm 0.0003$ &$0.8774$  \\
		\hline\\[-2.5ex]
		$\chi^2$ & 607  & 43.12\\
		$N_{\rm data}$ & 357 & 56 \\ 
           $\chi^2 /N_{\rm dof}$ & 1.699 & 0.77 \\
		\botrule
	\end{tabular}\label{tab:Arrington-FF}}
\end{table}

\begin{figure}[!]
  \begin{center}
    \leavevmode
    \epsfig{figure=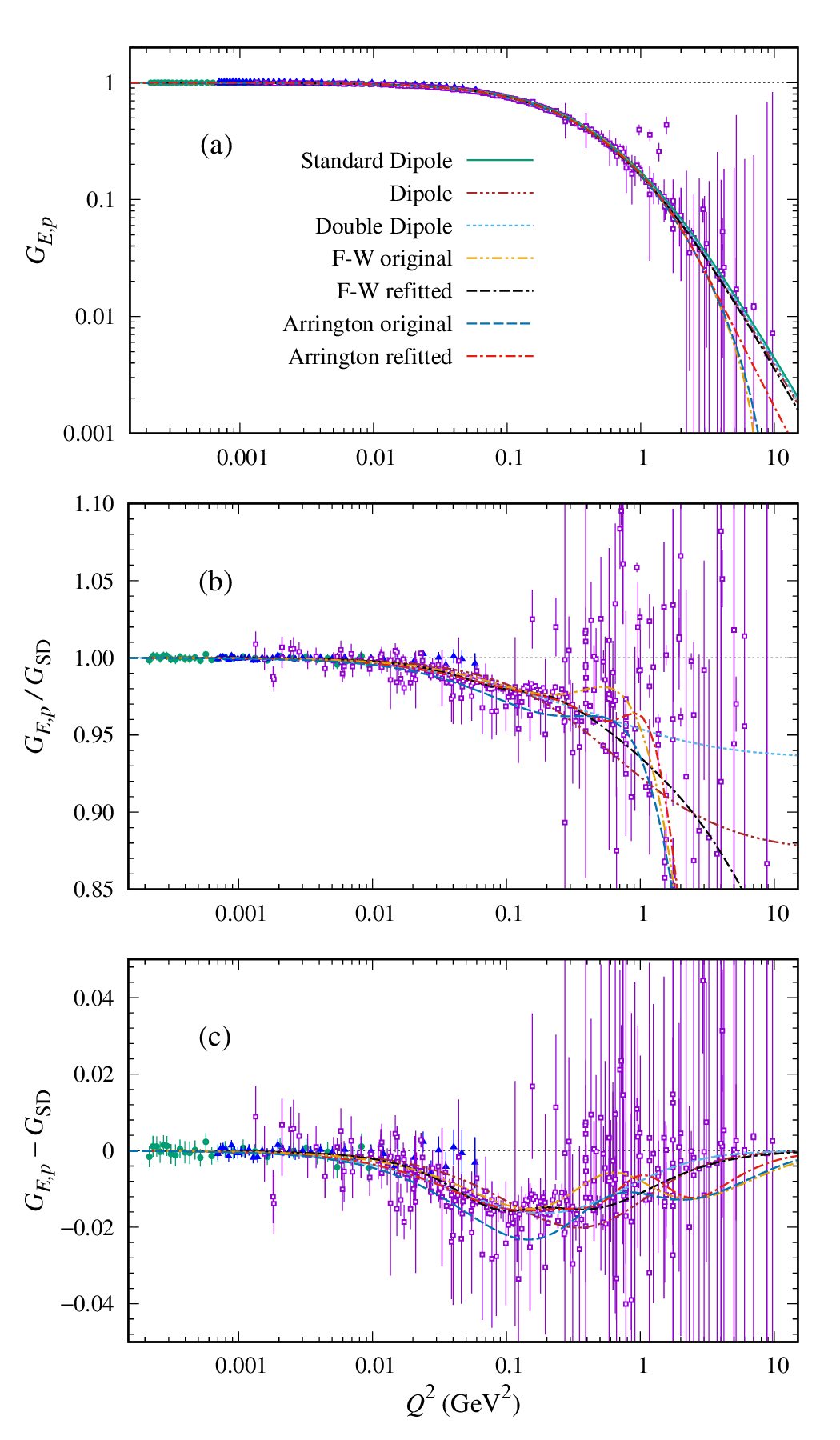,width=100mm}
    \caption{The proton electric form factor $G_{E,p}$  
      obtained from the present and previous works compared 
      with experimental data. The form factor is displayed
      in three different forms, i.e., (a) the original one
      $G_{E,p}$, (b) divided by the standard dipole one
      $G_{E,p}/G_{\rm SD}$, and (c) subtracted from the standard 
      dipole one $G_{E,p} - G_{\rm SD}$. The models of Friedrich-Walcher
      (F-W) and Arrington are obtained from 
      Refs.~\cite{Friedrich:2003iz} and \cite{Arrington:2007ux}.
      See Table \ref{exp_data} for the references of the experimental data.
      The new experimental data from the PRad collaboration 
      {\it et al.}\, \cite{Xiong:2019umf} are shown in solid green circles 
      (obtained with $E_e=1.1$ GeV)
      and solid blue triangles (obtained with $E_e=2.2$ GeV).
      }
   \label{fig:GEp} 
  \end{center}
\end{figure}

For the electric form factor $G_{E,p}$, comparisons between the fitted 
models and experimental data are shown in Fig.~\ref{fig:GEp}, where
we display the electric form factor $G_{E,p}$, the electric form factor
in the unit of the standard dipole one $G_{E,p}/G_{\rm SD}$, and 
the difference between the electric and the standard dipole form factors 
$G_{E,p}-G_{\rm SD}$, as functions of $Q^2$. These three 
methods of presenting the proton electric form factor are intended 
to look for the best and sensitive 
way to compare the performance of different parameterizations. 

From panel (a) of 
Fig.~\ref{fig:GEp} we can see that the plot of $G_{E,p}$ cannot clearly
distinguish the difference between the 
7 parameterizations discussed above. Only
at very high $Q^2$ the difference among them shows up. Unfortunately, the 
presently available data cannot single out the best parameterization
from our calculation. Moreover, the structure at $Q^2\approx 0.2$ GeV$^2$
cannot be seen here.

The electric form factor in the unit of the standard dipole one 
$G_{E,p}/G_{\rm SD}$ shown in panel (b) of Fig.~\ref{fig:GEp} starts to
display the difference between the parameterizations at $Q^2\gtrsim 0.2$ 
GeV$^2$. However, since the interesting structure is shown by the data
at $Q^2\approx 0.2$ GeV$^2$, the plot of $G_{E,p}/G_{\rm SD}$ does not
help too much in this case. Furthermore, as shown in this panel, relative 
to the standard dipole one, most of the parameterizations quickly 
decrease 
as the $Q^2$ increases from 1 GeV$^2$. Therefore, comparison between data
and models is less useful for the present purpose.

Only the third method, which uses 
$\Delta\equiv G_{E,p}-G_{\rm SD}$ shown in panel (c), 
gives the clearest information 
for our present purpose, since the structure around $0.2$ GeV$^2$ is clearly 
revealed by experimental data and model calculations. Interestingly, panel (c) 
of Fig.~\ref{fig:GEp} also indicates that the proton electric form factor 
$G_{E,p}$ is approaching the standard dipole one $G_{\rm SD}$ not only at low 
$Q^2$, but also at high $Q^2$ region, although in the latter experimental data
are wildly scattered with large error bars. The fact that all form factors
approaches the standard dipole one does not appear in panel (b) since all form
factors decrease faster than the standard dipole one. 
However, since at high
$Q^2$ all form factors are very tiny and inaccessible through experiment, we
may conclude that the third method is the best one to compare all form factor 
models. From panel (c) of Fig.~\ref{fig:GEp} it is clear that the structures 
near $0.2$ GeV$^2$ and the differences between the presented models could 
be investigated if we had experimental data with comparable accuracies to 
those at low $Q^2$.

\begin{figure}[t]
  \begin{center}
    \leavevmode
    \epsfig{figure=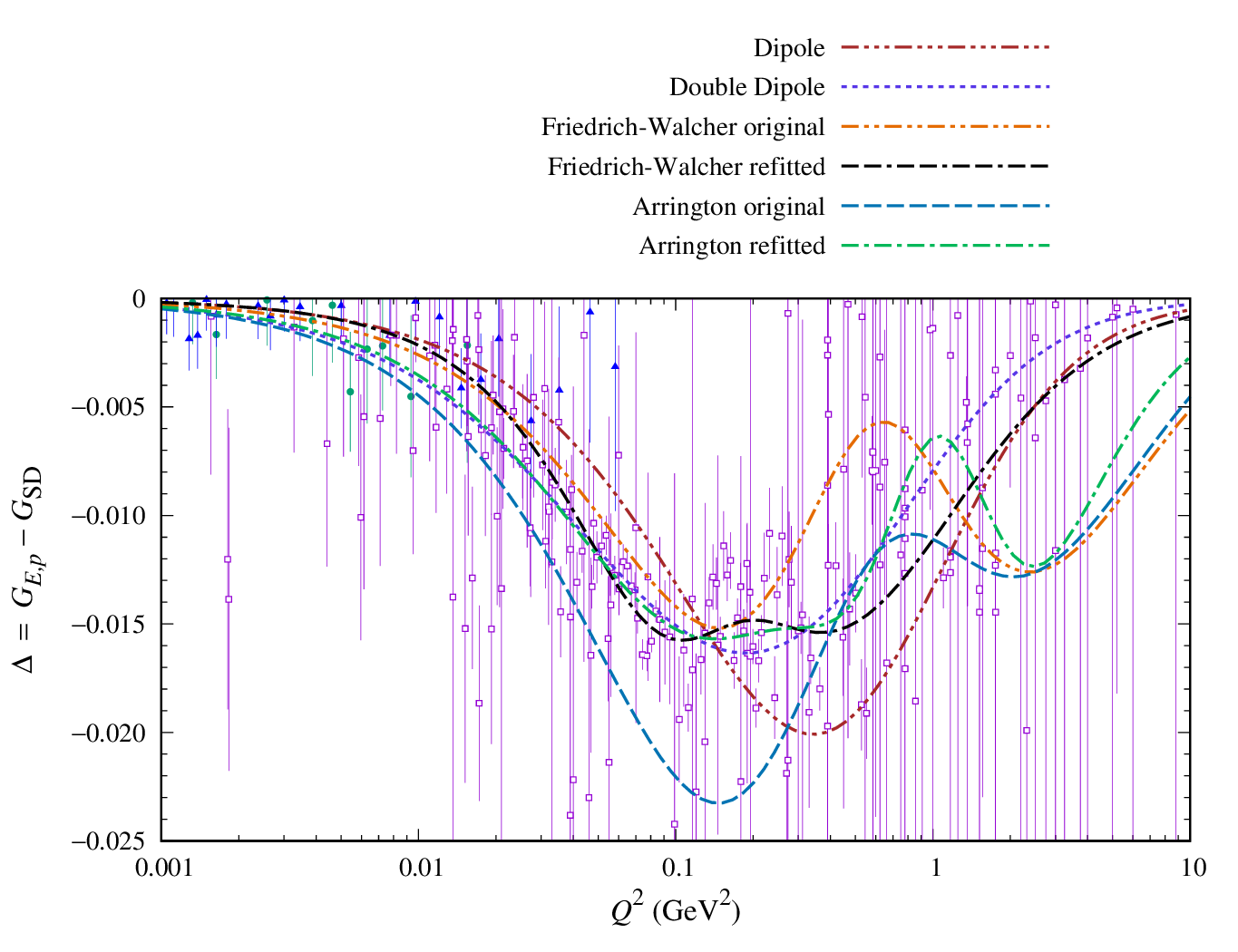,width=130mm}
    \caption{As in panel (c) of Fig.~\ref{fig:GEp}, but zoomed at the
      structure near $Q^2= 0.2$ GeV$^2$. 
      }
   \label{fig:gep_min_gsd} 
  \end{center}
\end{figure}

To investigate the structure near $Q^2= 0.2$ GeV$^2$ let us focus
our attention on panel (c) of Fig.~\ref{fig:GEp}, especially in the range of
$0.001 \lesssim Q^2\lesssim 10$ GeV$^2$. This is shown in Fig.~\ref{fig:gep_min_gsd},
where we compare the analyzed models with experimental data. By looking at the
data in the range of $0.05 \lesssim Q^2\lesssim 5$ GeV$^2$ we can infer that
there are at least two minima located at $Q^2\approx 0.1$ and 0.2 GeV$^2$. Another
possible minimum is seen at $Q^2\approx 0.4$ GeV$^2$, which is less visible than
the previous two. Clearly, these minima cannot be produced by the dipole
form factor, since as shown in Fig.~\ref{fig:gep_min_gsd},
this form factor can
only create one minimum, due to its smooth behavior. The same phenomenon is
also displayed by the double dipole one. 

As stated by Friedrich and Walcher \cite{Friedrich:2003iz},
the structure can be best described by
using the non-smooth term $G_{\rm ns}$ given by Eq.~(\ref{eqn:non-smooth_FW}).
As shown in Fig.~\ref{fig:gep_min_gsd}, by using 
this term, the original F-W
model is able to produce two minima at $Q^2\approx 0.15$ and 2 GeV$^2$. The 
refitted F-W model shifts these two minima closer to experimental data, i.e.,
at $Q^2\approx 0.1$ and 0.3 GeV$^2$. 

The original Arrington model also produces two minima with their positions very
close to those of the original F-W one. The two models differ only in their form factor
magnitudes. As in the case of refitted F-W model, the refitted Arrington model is
much closer to experimental data. The remarkable difference here is 
that the refitted
Arrington model produces three minima at $Q^2\approx 0.1, 0.3$ 
and 2 GeV$^2$. It is of course somewhat strange that both original F-W and 
Arrington models have a minimum at 2 GeV$^2$, whereas almost no experimental
data support this phenomenon. Even the refitted Arrington model has
also a minimum at this position. Presumably, it is the way of the models to
produce a peak at 0.8 GeV$^2$, where a number of experimental data points are
located. 

From Fig.~\ref{fig:gep_min_gsd} we observe that for $Q^2\to 0$ the 
softest slope (relative to the horizontal $\Delta=0$ line) 
is obtained for the Dipole and refitted F-W models, whereas the steepest one 
is given by the original Arrington model. From this 
observation, we can estimate
that the smallest radius is obtained by the Dipole and refitted F-W models, 
while the largest radius is obtained by the original Arrington model. This 
is proven in Fig.~\ref{fig:radius_el} and will be discussed later. 
At this stage it is important to emphasize that the different slopes 
obtained from different models shown in Fig.~\ref{fig:gep_min_gsd} 
are strongly related to different descriptions of the 
structure near $Q^2\approx 0.2$ GeV$^2$.
In other words, the experimental data at this kinematics
significantly influence the calculated proton radius. Therefore, our present 
work recommends that future experiments should focus on this kinematics
instead of going to lower $Q^2$. 
We note that similar conclusion can be also found in
previous works \cite{Pacetti:2018wwk,Hoballah:2018szw,Pacetti:2021fji}.
By comparing the models shown 
in Fig.~\ref{fig:gep_min_gsd}, we believe that experimental data with 
the same accuracy as the latest data from JLab, for instance, would 
significantly help to clarify whether the three minima really exist 
and, therefore, to single out the best parameterization
that globally fits the form factor data from 0.0001 to 10 GeV$^2$.

Now, let us consider the extracted proton radius from the different
models that we have discussed above. The result is shown in 
Fig.~\ref{fig:radius_el} by the open circles. There are many
analyses of this radius in the literature and, as a consequence, only
recent results are displayed in Fig.~\ref{fig:radius_el}.

A newer radius obtained 
by using initial state radiation (ISR) in MAMI Mainz and formally published in 
2021 is smaller than the Mainz 2010 result. However, the quoted error
is significantly large, making it consistent with both 
values obtained from the conventional elastic 
electron-proton scattering experiments 
and muonic hydrogen spectroscopic measurement.
On the other hand, instead of measuring the cross section, the JLab 
Hall A experiment \cite{Zhan:2011ji} utilized the polarization transfer of 
electron-proton elastic scattering in the region of $Q^2=0.3-0.7$ GeV$^2$ 
and obtained the radius which is consistent with the value 
obtained from the Mainz-2010 
electron-proton scattering experiment \cite{bernauer-2010}.

More recently, the PRad collaboration performed an experiment at JLab and
succeeded to reach much lower $Q^2$ by using a new electromagnetic calorimeter
that could cover a larger area, with a hole in the center to allow the 
non-scattered electrons to pass through, and therefore could reach 
much smaller scattering angles \cite{Xiong:2019umf}.
It is widely known that the problem to reach smaller $Q^2$ is related to
the problem of dealing with very forward scattering angles. The result of
this experiment is more precise than that of the polarization transfer
experiment at Hall A. As shown in Fig.~\ref{fig:radius_el}, a smaller
radius consistent with the value from 
muonic hydrogen spectroscopic measurement \cite{Pohl:2010zza}
was obtained. As discussed in the previous
Section we have refitted the PRad data by means of a dipole form factor and
obtained a consistent result as shown in the same figure. The smaller error
obtained in this case originates from the nature of dipole form factor,
which has only one parameter and, as a consequence, 
a very simple error formula given by
Eq.~(\ref{eq:error_rad_dipole}), where $\Delta\Lambda$ is obtained 
from {\small MINUIT}. This phenomenon was also found in our previous study 
\cite{Mart:2013gfa,Mart:2013gfa2,Mart:2013gfa3,Mart:makara}.

The next interesting result is shown by four theoretical calculations,
with two of them consistent with the 
muonic hydrogen value, whereas the other two consistent with 
the value from electron-proton scattering experiment.
By using an improved two-pion continuum,
Lin {\it et al}. found smaller radius \cite{Lin:2021umk} consistent 
with the muonic hydrogen value. A similar result was obtained by
Cui {\it et al.}, who analyzed the PRad data by means of a statistical
sampling approach. Different result was obtained by Arrington and Sick
\cite{Arrington:2015ria}. By using a global fit to elastic electron-proton
data they obtained a radius of $0.879\pm 0.011$ fm, which is 
identical to the CODATA-2014 value. Sick \cite{Sick:2018fzn} continued
the investigation by constraining the extrapolation to very low $Q^2$ 
region, where no experimental data are available, by using experimental 
data at finite $Q^2$ in order to reduce model dependence. He obtained
a radius value of $0.887\pm 0.012$ fm, which is consistent with the
value from electron-proton scattering experiment
too, as shown in Fig.~\ref{fig:radius_el}.

It is interesting to see that our fits (indicated by open circles in 
Fig.~\ref{fig:radius_el}) using dipole form factor to both low $Q^2$ 
and all $Q^2$ data are mostly closer to the value obtained 
from the muonic hydrogen atom spectroscopy. The relatively 
large difference obtained from fit to PRad and fit to all low $Q^2$ 
data is a clear indication of the difference between the two data sets,
as has been stated in the previous section. It is also interesting that,
by using all data, including the highest $Q^2$ ones, the extracted 
radius decreases and approaches the muonic hydrogen atom
value. This is the
clear evidence that the high $Q^2$ data have significant effect on
the extracted radius. As a consequence, high $Q^2$ data could become
a stringent constraint to the existing models that try to extract 
the proton charge radius.

Surprisingly, fitting to all data but with a double dipole form factor 
results in a larger radius, consistent with the value 
obtained from the conventional electron-proton scattering experiment.
Once
again, this result indicates that the extraction of radius also depends 
strongly on the form factor model. Compared to the dipole form factor, 
the larger error in this case originates from the larger number 
of degrees of freedom (free parameters). However, if we look back to
Fig.~\ref{fig:gep_min_gsd}, we observe that the double dipole form 
factor yields only one minimum at $Q^2\approx 0.2$ GeV$^2$ and, 
therefore, does not precisely reproduce the structure shown by
experimental data at this kinematics. 

As previously discussed, the more precise models that reproduce the
structure around $Q^2\approx 0.2$ GeV$^2$ are the F-W
and Arrington models. Original calculations of these models yield
also different radii. Whereas the original F-W model predicted a smaller
radius than the muonic hydrogen atom one, 
the Arrington model corroborates the conventional 
electron-proton scattering radius. 

Interestingly, refitting the two models to new data shifts their radii 
closer to each other, i.e., increases the F-W radius and decreases the 
Arrington one. The origin of this phenomenon can be traced back to 
Fig.~\ref{fig:gep_min_gsd}. Refitting the F-W model shifts the second
minimum to a lower $Q^2$, where more experimental data show a minimum.
This process makes the corresponding slope to the $Q^2=0$ point 
steeper and, as a consequence, makes the extracted radius larger. 
In the case of Arrington model, the refitting process increases the 
number of minima from two to three. This is possible since, 
as shown in Eq.~(\ref{eq:arrington1}), 
the Arrington model has more complex
functions of $Q^2$. In this case, the new minimum is located at 
$Q^2\approx 0.4$ GeV$^2$. This new minimum makes the slope of form factor,
relative to the standard dipole one, 
softer and, therefore, makes the extracted radius smaller.

Note that as shown in Tables \ref{tab:FW-FF} and \ref{tab:Arrington-FF},
the difference between the $\chi^2$'s obtained 
from F-W and Arrington models 
is almost negligible (608 and 607). However, by carefully looking at
Fig.~\ref{fig:gep_min_gsd}, we believe that the F-W model is closer to
reproduce experimental data at two minima near 0.1 and 0.4 GeV$^2$,
whereas the
third minimum of Arrington model near 2 GeV$^2$ seems to be 
unsupported by experimental data. Therefore, our present study prefers
the refitted Friedrich-Walcher model that yields a radius of
$0.836\pm 0.011$ fm, consistent with the muonic hydrogen atom
value. Nevertheless, a more quantitative conclusion should wait for
more precise experimental data in this kinematical region. As stated
before, experimental data with accuracy comparable to the latest data
in the low $Q^2$ region would be very decisive to this end.

\begin{figure}[!]
  \begin{center}
    \leavevmode
    \epsfig{figure=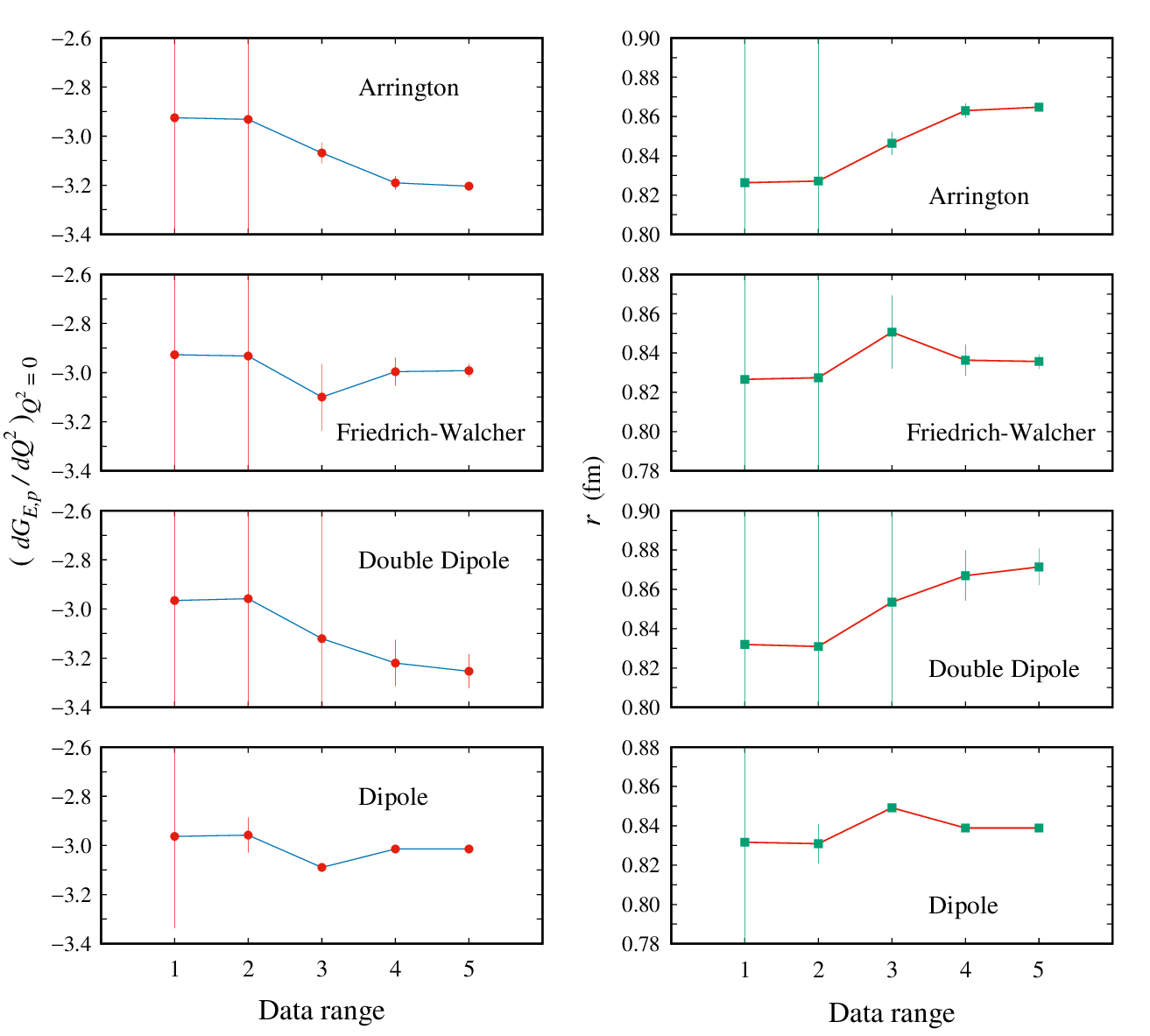,width=130mm}
    \caption{Derivatives of the proton electric form 
      factor calculated at photon point 
      $\left|dG_{E,p}/dQ^2\right|_{Q^2=0}$ (left panels) and 
      the corresponding proton charge radii (right panels) as
      functions of the data range used in the fitting database
      for different types of form factor indicated in each panel. 
      The data ranges are (1) $0<Q^2<10^{-3}$, (2) $0<Q^2<10^{-2}$, 
      (3) $0<Q^2<10^{-1}$, (4) $0<Q^2<1$, and (5) $0<Q^2<10$ GeV$^2$.
      }
   \label{fig:derivative}
  \end{center}
\end{figure}

To emphasize the importance of including higher $Q^2$ data in the
extraction of proton charge radius, in Fig.~\ref{fig:derivative} 
we show the effect of the data range used in the fit on the 
extracted form factor derivatives at photon point along with 
the corresponding proton charge radii. For this purpose we divide 
the available data into five data ranges written in detail 
in the caption of Fig.~\ref{fig:derivative}.

From the data shown in Fig.~\ref{fig:GEp}(c) we would expect that 
the largest effect could be obtained if we included the data near the 
structure, i.e., with $Q^2$ between 0.01 and 1 GeV$^2$, which corresponds 
to the use of the data ranges 3 and 4. This is proven by both the form 
factor derivatives at photon point and  the corresponding proton 
charge radii shown in Fig.~\ref{fig:derivative}. All form factors
show significant derivative or radius changes once we switch from 
the data range 2 to 3, as well as from the data range
3 to 4. Furthermore, as expected from the data
shown in Fig.~\ref{fig:GEp}(c), the use of data range 1 or 2 does not
produce different derivatives or radii. The same phenomenon is also
observed in the case of data range 4 or 5. This result emphasizes
our previous finding that the inclusion of data with $Q^2$ between 
0.01 and 1 GeV$^2$ significantly changes the form factor slope to 
the $Q^2=0$ point, as discussed above. 

The large error bars obtained from fitting the data range 1 or 2,
especially in the case of Arrington and Friedrich-Walcher 
form factors, originate from the number of used free parameters, 
which is unnecessarily too large to fit the relatively few 
data points. This is not the case for the dipole form factor, 
since it has only one free parameter.
To conclude this section, we have shown that the use of higher $Q^2$
data has a significant effect on the extracted proton charge radius.

\subsection{The magnetic radius of proton}

\begin{figure}[!]
  \begin{center}
    \leavevmode
    \epsfig{figure=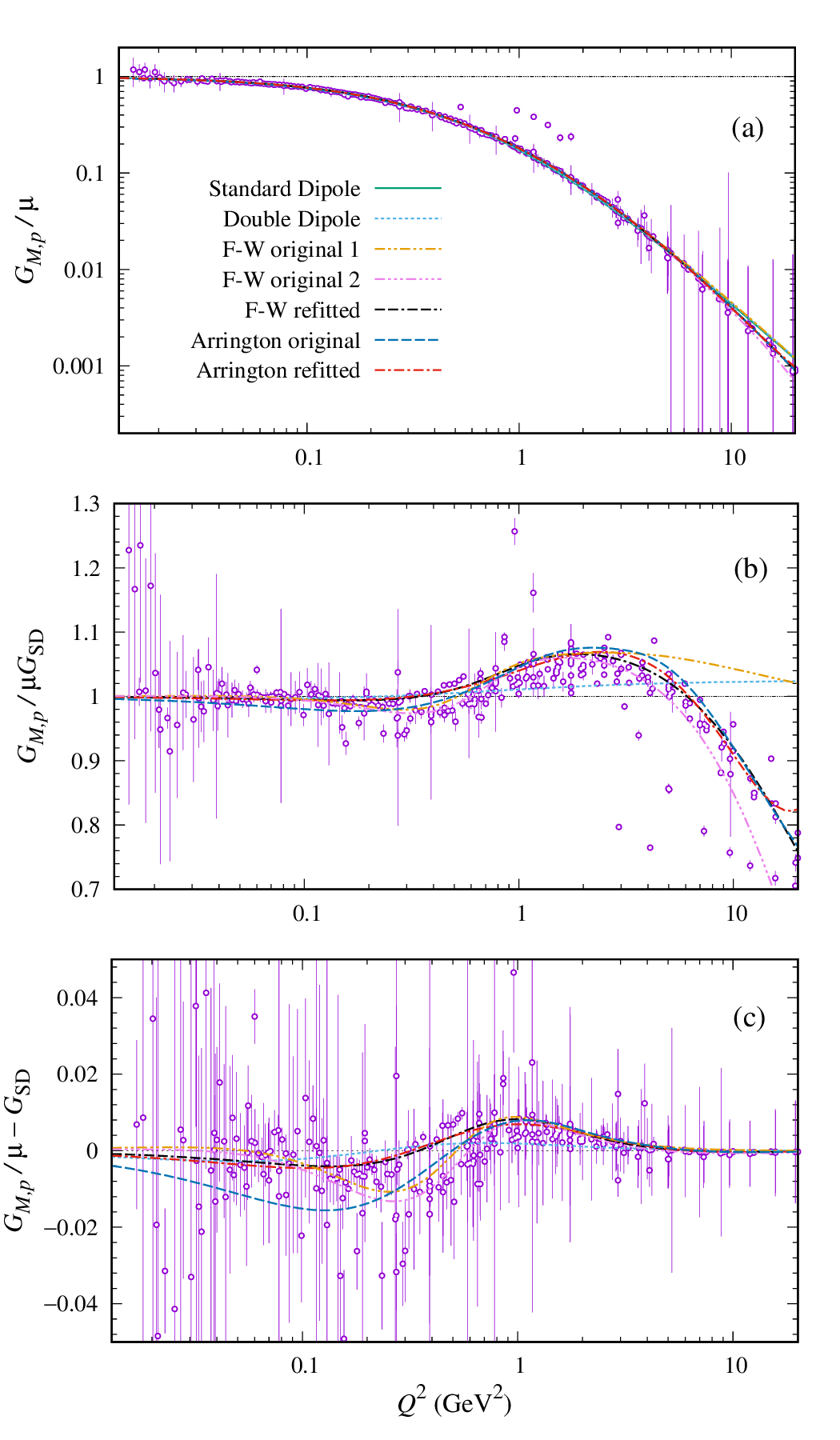,width=100mm}
    \caption{The proton magnetic form factor $G_{M,p}/\mu$  
      obtained from the present and previous works compared 
      with experimental data. The form factor is displayed
      in three different forms, i.e., (a) the original one
      $G_{M,p}/\mu$, (b) divided by the standard dipole one
      $G_{M,p}/\mu G_{\rm SD}$, and (c) subtracted from the standard 
      dipole one $G_{M,p}/\mu - G_{\rm SD}$, with $\mu$ the proton
      magnetic moment in the unit of nuclear magneton. 
      As in Fig.~\ref{fig:GEp}
      the models of Friedrich-Walcher
      (F-W) and Arrington are obtained from 
      Refs.~\cite{Friedrich:2003iz} and~ \cite{Arrington:2007ux}{.}
}
   \label{fig:GMp} 
  \end{center}
\end{figure}

\begin{figure}[t]
  \begin{center}
    \leavevmode
    \epsfig{figure=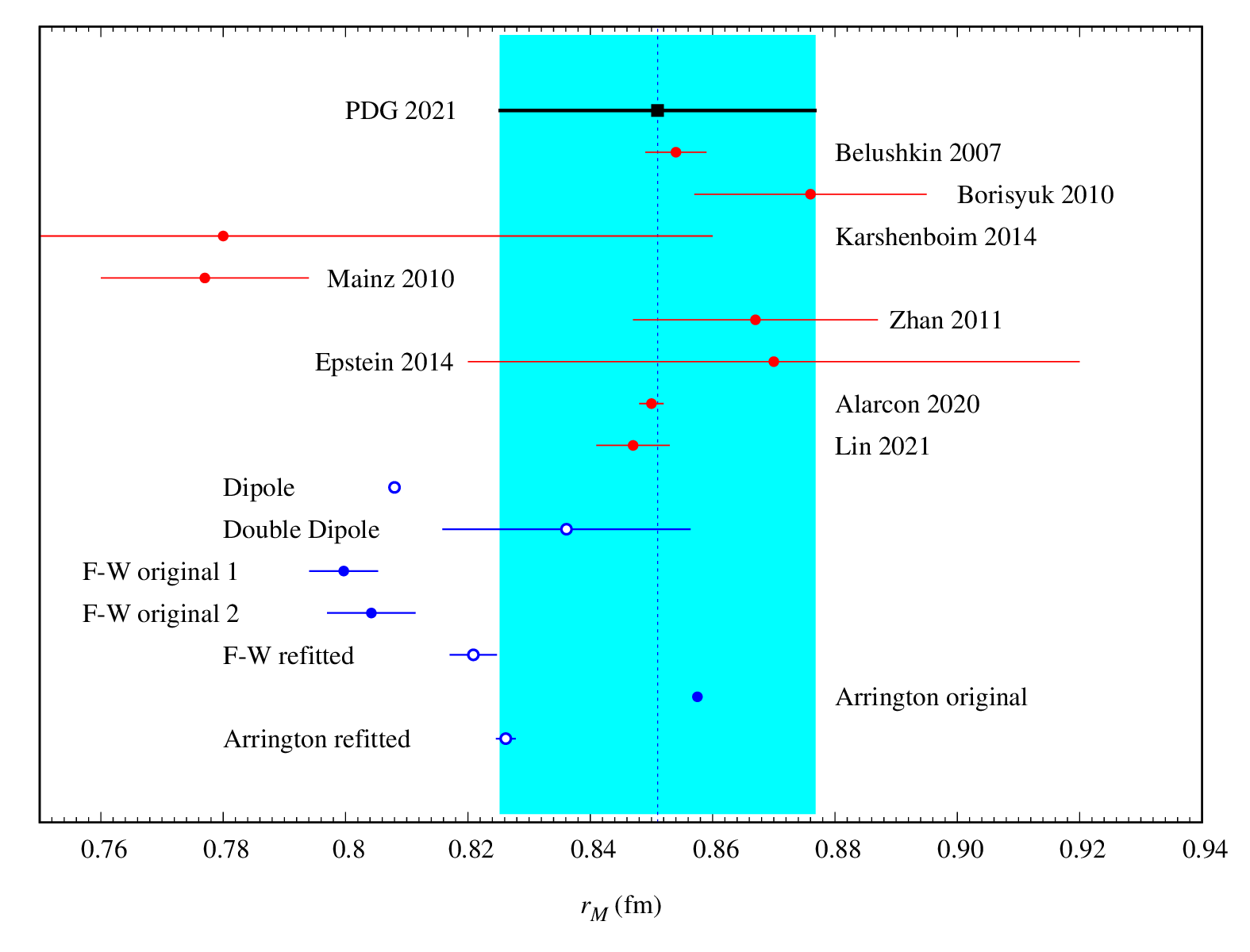,width=130mm}
    \caption{As in Fig.~\ref{fig:radius_el} but for 
      the proton magnetic radius 
      $r_{M}\equiv \sqrt{\langle r_{M,p}^2 \rangle}$ obtained
      from the present and previous works compared with the
      recommended value obtained from PDG 2021 \cite{pdg2021}.
      The previous works are obtained from theoretical works of 
      Belushkin 2007 \cite{Belushkin:2006qa}, 
      Borisyuk 2010 \cite{Borisyuk:2009mg}, 
      Zhan 2011 \cite{Zhan:2011ji}, 
      Karshenboim 2014 \cite{Karshenboim:2014vea}, 
      Epstein 2014 \cite{Epstein:2014zua},
      Alarcon 2020 \cite{Alarcon:2020kcz}, Lin 2021 \cite{Lin:2021umk},
      and experimental work from Mainz 2010 \cite{A1:2010nsl}.
      As in Fig.~\ref{fig:radius_el}, the radius error bar of the 
      Arrington original
      model was not reported.}
   \label{fig:radius_mag} 
  \end{center}
\end{figure}

For completeness, in Fig.~\ref{fig:GMp} we also present the proton
magnetic form factor $G_{M,p}/\mu$, where we display the
behavior of the form factor in the form of $G_{M,p}/\mu G_{\rm SD}$ and
$G_{M,p}/\mu - G_{\rm SD}$, similar to the case of electric form factor
discussed above. It is important to note here that the ratio
$G_{M,p}/\mu G_{\rm SD}$ reveals the second structure (peak or maximum) 
clearly at $Q^2\approx 2$ GeV$^2$ while the difference 
$G_{M,p}/\mu - G_{\rm SD}$ display both structures at $Q^2\approx 0.2$ 
and 1-2 GeV$^2$ moderately. 

Different from the electric form factor, here we see that the refitted
F-W and Arrington models agree with each other, while the difference 
between the original models is more obvious in panel (c) of 
Fig.~\ref{fig:GMp}, since the difference only exists in the first
structure (minimum) at 0.2 GeV$^2$. 

The extracted proton magnetic radii $r_{M}$ obtained 
in the present work are shown in 
Fig.~\ref{fig:radius_mag} by open circles, along with those previously 
calculated and available in literature. In this figure we compare the radii
with that of the Particle Data Group \cite{pdg2021} (PDG),
i.e., $r_M=0.851\pm 0.026$ fm. Here we
can see that most of the calculations agree with the PDG estimate. 
This is very different from the case of charge radius. 
The proton magnetic radii extracted from refitting the F-W and Arrington models
are also moving closer to each other, as expected from the behavior of 
their form factors for small $Q^2$ shown in Fig.~\ref{fig:GMp}. While the
refitted F-W model yields $r_M=0.821\pm 0.004$ fm, the Arrington refitted one
yields $r_M=0.826\pm 0.002$ fm. Thus, the extracted 
radii from the two models are consistent to each other as well as to the
PDG estimate. The present calculation indicates that refitting models with
the new data from very low to very high $Q^2$ can improve the agreement
of the extracted proton magnetic radii.

\section{Summary and Conclusion}
\label{sec:conclusion}
We have revisited the extraction of proton electric and magnetic radii.
To this end we presented a number of form factor models 
commonly used to extract the radii. The formulation of the 
models along with the
corresponding error were also derived. Special attention was given 
to the higher $Q^2$ region, where two structures are exhibited by 
experimental data. To this end we exploited two form factor models that were 
previously used to globally fit the data, i.e., the 
Friedrich-Walcher and Arrington parameterizations, along with the dipole
and double-dipole ones. 
Since the new
PRad data at very low $Q^2$ gave surprisingly small charge radius, 
consistent with the value from muonic hydrogen 
spectroscopic measurement, but in contrast to the
previous experiments using elastic electron-proton scattering, we first 
reanalyzed the data by using a simple dipole form factor. We extracted
the radius and obtain a relatively small value, consistent with the PRad's 
reported value, although we used different form factor. 
By including the higher $Q^2$ data we observed at least two structures
(minima) at $Q^2\approx 0.1$ and 0.2 GeV$^2$. The dipole and double dipole
form factors produced only one minimum. The refitted Friedrich-Walcher and
Arrington form factors could nicely reproduce these structures, but the
Friedrich-Walcher one seems to be the more natural. 
The latter yielded a radius of $0.836\pm 0.011$ fm, which is consistent 
to the radius obtained from muonic hydrogen atom. 
We have also reanalyzed the magnetic
form factor data and obtained the magnetic radii consistent with the
PDG 2021 estimate if we used the Friedrich-Walcher and Arrington form
factors. Although our analysis indicates that the higher $Q^2$ data play a 
significant role in extracting the proton electromagnetic radii, a more
definite conclusion could be drawn only after we have new data at this 
kinematics with error comparable to the present experiments.

\section*{Acknowledgment}
This work was supported by a special grant provided 
by Universitas Indonesia 
under contract No. NKB-587/UN2.RST/HKP.05.00/2021.

\end{document}